\documentclass[twocolumn,twoside,superscriptaddress,pra,aps,showpacs]{revtex4}
\usepackage{amsmath,bm}
\usepackage{graphicx}
\usepackage{psfrag}                  % manages figures
\usepackage{epsfig}                 % manages figures
\usepackage{float}
\usepackage{xspace}                  
\usepackage{hyperref}
\usepackage{color}

\usepackage{algorithm}

\usepackage{algpseudocode}

\begin{document}

\title{Practical long-distance quantum communication \\ using concatenated entanglement swapping}

\author{Aeysha Khalique}
\affiliation{Institute for Quantum Science and Technology, University of Calgary, Alberta T2N 1N4, Canada}
\affiliation{Centre for Advanced Mathematics and Physics, National University of Sciences and Technology, H-12 Islamabad, Pakistan}
\author{Wolfgang Tittel}
\affiliation{Institute for Quantum Science and Technology and Department of Physics and Astronomy, University of Calgary, Alberta T2N 1N4, Canada}
\author{Barry C. Sanders}
\affiliation{Institute for Quantum Science and Technology, University of Calgary, Alberta T2N 1N4, Canada}

\date{\today}

\begin{abstract}
We construct a theory for long-distance quantum communication based on sharing entanglement 
through a linear chain of $N$ elementary swapping segments of  length~$L=Nl$ where $l$ is the length of each elementary swap setup.
Entanglement swapping is achieved by linear optics, photon counting and post-selection,
and we include effects due to multi-photon sources, transmission loss and detector inefficiencies and dark counts.
Specifically we calculate the resultant four-mode state shared by the two parties at the two ends of the chain,
and we derive the two-photon coincidence rate expected for this state
and thereby the visibility
of this long-range entangled state.
The expression is a nested sum with each sum extending from zero to infinite photons,
and we solve the case $N=2$ exactly for the ideal case (zero dark counts, unit-efficiency detectors and no transmission loss)
and numerically for $N=2$ in the non-ideal case with truncation at $n_\text{max}=3$ photons in each mode.
For the general case, we show that the 
computational complexity for the numerical solution is $n_\text{max}^{12N}$.
\end{abstract}

\maketitle

\section{Introduction}

In practice long-distance quantum communication based on transmission of qubits suffers from a bound on transmission length because qubits can get lost along the way.
Detector dark counts further complicate matters by allowing detectors to register a spurious count even if the original qubit is lost, and the combination of loss and dark counts
limits the distance to around 200~km~\cite{SWV+09,CWL+10}. Quantum repeaters provide a means to overcome
this problem with a resource overhead that is at worst a polynomial function of the desired transmission distance~$L$~\cite{DLC+01}.
However, quantum repeaters are still in an early stage or research~\cite{SSR+11,MRS12,ABB+12}.
On the other hand, entanglement swapping~\cite{PBW+98} can be used to extend the distance for quantum communication
with current technology albeit with a resource overhead that is exponential in~$L$~\cite{GRT+02}.
This set-up is known as a `quantum relay'.
Although an exponential overhead is daunting, it is still better than a distance bound that renders
quantum communication at distances greater than this bound impossible.

Here we consider the practical quantum relay where we explicitly consider multi-photon events, transmission loss
and detector inefficiencies and dark counts. 
Our aim is to construct a formal theory for a linear chain of $N$ elementary swapping segments subject to multi-photon events and 
detector imperfections, with this theory delivering an expression for the resultant entangled four-mode 
state at the ends of the chain post-selected on detection records at entanglement swapping devices along the chain.
This theory delivers not only the resultant state but also the two-photon coincidence rate from
which the two-photon visibility of the entangled state is calculated.
Our results show that multiphoton effects are an important deleterious contributor to two-photon visibility
for a long-distance array of concatenated entanglement swapping.

Our theory requires that $N=2^\imath$ for $\imath$ some positive integer due to symmetry in the calculations,
and we solve the expression exactly for $N=2$ in the ideal case
and numerically for the non-ideal case and $N=2$.
Previously only the $N=1$ (equivalent to one swap) case has been studied under practical conditions~\cite{SHS+09},
and our work extends that work but in a nontrivial way.
In particular our calculations only work for $N=2, 4, 8, \dots$,
with this restriction to make the calculations easier to perform,
whereas the previous result effectively considers only $\imath=0$.
Entanglement swapping concatenation has been studied before but without multiphoton effects~\cite{CGR05}.

We determine the computational complexity for the numerical simulation as a function of~$N$
and the truncation~$n_\text{max}$ of photon number in each mode.
Specifically the algorithm that we developed is inefficient as its runtime scales as $n_\text{max}^{12N}$.
By truncating at $n_\text{max}=3$ we are able to solve for $N=2$ using a 
message-passing interface parallel program on a supercomputer.
For a fixed set of parameters and with truncation to a maximum of three photons for each of the 16 modes, the Fortran program running at 2.66~GHz on an Intel Xeon E5430 quad-core processor with 8~GB of memory required approximately six hours to compute two-photon coincidence probability on a single core. Hence code parallelization was necessary to deliver probabilities for wide range of inputs in reasonable time, by making use of multiple cores. The given inputs included parametric down conversion pump rate~$\chi$, dark-count rate~$\wp$, detector efficiency~$\eta$, transmission loss
and polarization rotator angles~$\tilde{\alpha}$ for Alice on the left end of the chain of entanglement swappings
and~$\tilde{\delta}$ for Bob on the right end.

Our paper proceeds as follows.
In Sec.~\ref{sec:background} we provide a background on practical entanglement swapping.
This background concerns only the $N=1$ case
and introduces the concepts required for subsequent analysis.
The $N=2$ case is developed in Sec.~\ref{sec:N=2}
including computing the conditioned entangled state shared between Alice on the left and Bob on the right
and the resultant two-photon coincidence probability and visibility.
In Sec.~\ref{sec:N} we derive the general state and visibility formul\ae\ for arbitrary number~$N$ concatenations of 
entanglement swapping.
We discuss the complexity for numerically solving the general case in this section as well.
In Sec.~\ref{sec:conclusions} we summarize the result and present our conclusions.

\section{Background: practical entanglement swapping}
\label{sec:background}

\begin{figure}
\includegraphics[scale=0.5]{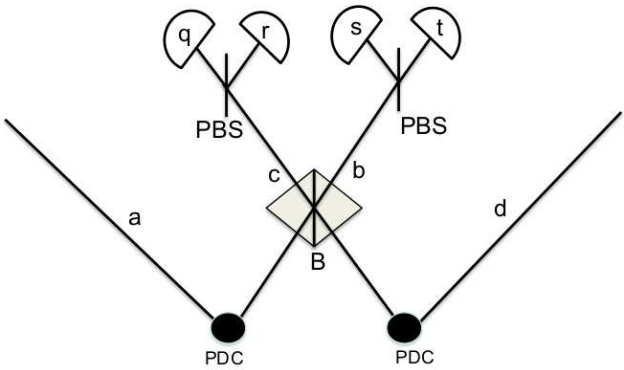}
\caption{
	$N=1$ entanglement swapping.
	Entangled states are prepared in each pair of modes by two parametric down converters (PDCs).
	Bell measurement on spatial modes~$b$ and~$c$
	are used to conditionally prepare an entangled state between modes~$a$ and~$d$.
	Bell measurement comprises of combining modes~$b$ and~$c$ on beam splitter~B 
	followed by polarization separation at one polarizing beam splitter (PBS) for each mode.
	The resultant photon counts at each of the four detectors
	(the `detector four tuple') is $\{q,r,s,t\}$.
	}
\label{fig:OneSwap}
\end{figure}

In this section we reprise the case of a single entanglement swapping ($N=1$)
under practical conditions~\cite{SHS+09}
as these results inform us in how to solve the case of $N=2^\imath$, 
for $\imath$ any positive integer,
in subsequent sections.
The setup is shown in Fig.~\ref{fig:OneSwap}.
For simplicity we assume that qubits are encoded into polarization states. However our formalism is general and can also be applied to other realization of qubits e.g., time bin qubits. Two parametric downconversion (PDC) sources produce two-mode entangled states.
Ideally each of these two-mode entangled states corresponds to an entangled pair of photons,
but the realistic case involves the vacuum state~$|\text{vac}\rangle$ and higher-order Fock states.
The `right' mode from the `left' PDC and the `left' mode of the `right' PDC are subjected to joint Bell-state measurements.
The Bell-state measurement conditions the resultant state shared by modes~$a$ and~$d$,
which should then be in an entangled state despite the two modes~$a$ and~$d$ not being
entangled initially, hence the term ``entanglement swapping''.

Assuming that PDC generates a pure state,
the quantum state prepared by the two PDC sources is given as
\begin{align}
\label{eq:2PDC}
	\left|\chi\right\rangle
		=&\exp[i\chi(\hat{a}^\dagger_\text{H}\hat{b}^\dagger_\text{H}
			+\hat{a}_\text{H}\hat{b}_\text{H})\otimes
			\exp[i\chi(\hat{a}^\dagger_\text{V}\hat{b}^\dagger_\text{V}+\hat{a}_\text{V}\hat{b}_\text{V})
				\nonumber\\&
		\otimes\exp[i\chi(\hat{c}^\dagger_\text{H}\hat{d}^\dagger_\text{H}+\hat{c}_\text{H}\hat{d}_\text{H})
				\nonumber\\&
                   \otimes\exp[i\chi(\hat{c}^\dagger_\text{V}\hat{d}^\dagger_\text{V}+\hat{c}_\text{V}\hat{d}_\text{V})
                      \left|\text{vac}\right\rangle,
\end{align}
and the mixed-state case is readily generalized by making an incoherent mixture of the pure states~(\ref{eq:2PDC}).
Given that four imperfect detectors (a detector fourtuple)
(i.e., detectors having non-unit efficiency~$\eta$ and non-zero dark count rates~$\wp$)
in the Bell measurement yield readout $\{q,r,s,t\}$,
as explained in Fig.~\ref{fig:OneSwap},
the posteriori conditional probability for any readout $(ijkl)$
that four ideal detectors would have yielded, is
\begin{align}
	P^{qrst}_{ijkl}:=&p(ijkl|qrst)\nonumber\\
	\equiv&\frac{p(qrst|ijkl)p(ijkl)}{\sum\limits_{i',j',k',l'=0}^{\infty}p(qrst|i'j'k'l')p(i'j'k'l')}.
\end{align}
Note that as in \cite{SHS+09}, we include all transmission loss into the detector efficiency. As the detectors are independent,
\begin{equation}
	P(qrst|ijkl)=p(q|i)p(r|j)p(s|k)p(t|l).
\end{equation}

For photon number discriminating detector with efficiency~$\eta$ and dark count probability $\wp$,
\begin{equation}
	p(q|i)=\frac{(1-\eta)(1-\wp)}{1-\eta(1-\wp)}\left(\frac{\eta}{1-\eta}\right)^q(1-\eta)^i
                           G(i,q;\eta,\wp)
\end{equation}
for $i\ge q$
and
\begin{equation}
	p(q|i)=\frac{(1-\eta)(1-\wp)}{1-\eta(1-\wp)}\left[\frac{1-\eta}{\eta}b(\eta,\wp)\right] ^{q-i}                
                            \eta^iG(i,q;\eta,\wp)
\end{equation}
for $q>i$
where
\begin{equation}
	b(\eta,\wp):=\left[1+\frac{1-\eta}{\eta \wp}\right]^{-1}.
\end{equation}
Also
\begin{align}
G(\kappa,\lambda;\eta,\wp)
                =&\sum_{n=0}^\infty{\kappa\choose\lambda}          
                     {\kappa-\lambda+n\choose\kappa-\lambda}
                      \left[b(\eta,\wp)\right]^n\nonumber\\
                &\times\left[{}_2F_1\left(-n,-\lambda;\kappa-\lambda+1;\frac{\eta-1}
                     {\eta}\right)\right]^2
\end{align}
for $\kappa\ge\lambda$
and $G(\kappa,\lambda;\eta,\wp):=0$ for $\kappa<\lambda$.
In the above equations, ${_2F_1}$ is the hypergeometric function.

For threshold detector there are two possibilities, click or no-click, which result in
\begin{align}
	p(\textrm{no click}|i)
		=&p(q=0|i)
			\nonumber\\
	=&(1-\wp)[1-\eta(1-\wp)]^i
\end{align}
\begin{align}
	p(\textrm{click}|i)
		=&1-p(\textrm{no click}|i)
				\nonumber\\
		=&1-(1-\wp)[1-\eta(1-\wp)]^i
\end{align}

The state of the remaining modes $a$ and $d$ after recording photons counts $\{q,r,s,t\}$
is the mixed state
\begin{equation}
	\rho^{qrst}=\sum_{i,j,k,l}P^{qrst}_{ijkl}|\Phi_{ijkl}\rangle\langle\Phi_{ijkl}|,
\end{equation}
with $|\Phi_{ijkl}\rangle$ the state corresponding to the count $\{i,j,k,l\}$ for perfect detectors $\{b_H,b_V,c_V,c_H\}$.
This ideal state is
\begin{align}
\left|\Phi_{ijkl}\right\rangle
         =&\frac{1}{\sqrt{2^{i+j+k+l}i!j!k!l!}}
                 \sum_{\mu=0}^i\sum_{\nu=0}^j\sum_{\kappa=0}^k\sum_{\lambda=0}^l
                   (-1)^{\mu+\nu}
                   	\nonumber\\&\times
           {i\choose \mu}{j\choose \nu}{k\choose \kappa}{l\choose \lambda}
           \hat{a}_\text{H}^{\dagger\mu+\lambda}
                   	\nonumber\\&\times
           \hat{a}_\text{V}^{\dagger\nu+\kappa}
             \hat{d}_\text{H}^{\dagger i+l-\mu-\lambda}\hat{d}_\text{V}^{\dagger j+k-\nu-\kappa}|     
               \text{vac}\rangle.
\label{eq:statePhi}
\end{align}
The visibility of the modes~$a$ and~$d$ is calculated by passing these two modes through polarizer rotators and measuring the coincidence count at the detectors after passing through polarizer beam splitters. This procedure reflects the standard experimental approach.

\section{Concatenating two entanglement swappings}
\label{sec:N=2}

In this section we develop the theory of concatenated entanglement swapping under practical conditions.
The concept of concatenated entanglement swapping is that two or more entanglement swapping processes are 
combined into a single entanglement-swapping procedure.
In this section we only deal with the case of $N=2$ concatenated elementary entanglement swappings, which 
we are able to solve approximately in the numerical case.
Beyond $N=2$ is the subject of Sec.~\ref{sec:N} where we provide the formalism but do not solve numerically.

\subsection{Conditionally prepared state}
\label{subsec:conditional}

\begin{figure}
\includegraphics[scale=0.5]{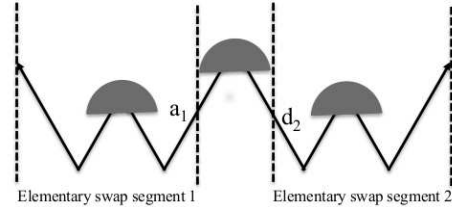}
\caption{
	Concatenation of $N=2$ elementary segments comprising of entanglement swapping setup: The cap on the inner arms represents a Bell-state measurement.
	Quantum communication distance is doubled by concatenating two elementary swaps by means of a third swap.
	Modes~$a_1$ and~$d_2$ from the 1st and 2nd swap are subjected to a Bell-state measurement
	to swap the entanglement to the right and left most arms.
	}
\label{fig:TwoSwaps}
\end{figure}

In Fig.~\ref{fig:TwoSwaps} we depict the case of $N=2$ concatenated entanglement swapping.
For the two elementary swaps shown in Fig.~\ref{fig:TwoSwaps}, we take the states of swap 1 and 2 as in
Eq.~(\ref{eq:statePhi}).
The modes~$a_1$ and~$d_2$ are combined on the $3^\text{rd}$ Bell-state measurement setup 
comprising a beam splitter $U_B$, polarizing beam splitters and a fourtuple of detectors. 
The operators $\hat{a}_{1,\text{H}}$ and $\hat{d}_{2,\text{H}}$  transform under the action of the beam splitter as
\begin{equation}
	U_B^\dagger\hat{a}_{1,\text{H}}^\dagger U_B
		=\frac{1}{\sqrt 2}\left(\hat{a}_{1,\text{H}}^\dagger-\hat{d}_{2,\text{H}}^\dagger\right)
\end{equation}
and
\begin{equation}
	U_B^\dagger\hat{d}_{2,\text{H}}^\dagger U_B
		=\frac{1}{\sqrt 2}\left(\hat{a}_{1,\text{H}}^\dagger+\hat{d}_{2,\text{H}}^\dagger\right),
\end{equation}
and $\hat{a}_{1,\text{V}}^\dagger$  and $\hat{d}_{2,\text{V}}^\dagger$ transform analogously.

Taking the readout at the detectors for the two horizontal modes~$a_{1,\text{H}}$ and~$d_{2,\text{H}}$
to be $\left|i_3,l_3\right\rangle$, we obtain
\begin{align}
	\left\langle i_3,l_3\right|&U_B \hat{a}_{1,\text{H}}^{\dagger\mu_1+\lambda_1}\hat{d}_{2,\text{H}}^{\dagger i_2+j_2-\mu_2-\lambda_2}\left|\text{vac}\right\rangle
			\nonumber\\
		=&\sum_{\gamma=0}^{\mu_1+\lambda_1}{{\mu_1+\lambda_1}\choose{\gamma}}{{i_3+l_3-\mu_1-\lambda_1}\choose{i_{3}-\gamma}}
			\nonumber\\&\times
	(-1)^{\mu_1+\lambda_1-\gamma}\sqrt{\frac{i_{3}!l_{3}!}{(2)^{i_{3}+l_{3}}}}
				\nonumber\\&\times
	\delta_{i_3+l_3,\mu_1+\lambda_1+i_2+l_2-\mu_2-\lambda_2}.
\label{eq:trans}
\end{align}

The transformation for vertical modes is similar.
Using Eq.~(\ref{eq:trans})
and taking the readout at the $3^\text{rd}$ fourtuple of detectors as $\{i_3,j_3,k_3,l_3\}$,
the unnormalized state of the remaining modes~$d_1$ and~$a_2$ is
\begin{widetext} 
\begin{align}
	\left|\Phi'_{\bm{ijkl}}\right\rangle=&\left\langle {i_{3}j_{3}k_{3}l_{3}}|U_B|\Phi_{i_1j_1k_1l_1}\Phi_{i_2j_2k_2l_2}\right\rangle\nonumber\\
	=&\prod_{p=1}^{2}\frac{1}{(\sqrt{2})^{i_p+j_p+k_p+l_p}\sqrt{i_p!j_p!k_p!l_p!}}
		\frac{(\tanh\chi)^{i_p+j_p+k_p+l_p}}{\cosh^8\chi}
			%\nonumber\\&\times
\sum_{\mu_p=0}^{i_p}\sum_{\nu_p=0}^{j_p}\sum_{\kappa_p=0}^{k_p}\sum_{\lambda_p=0}^{l_p}
(-1)^{\mu_p+\nu_p}{i_p\choose \mu_p}{j_p\choose \nu_p}{k_p\choose \kappa_p}{l_p\choose \lambda_p}\nonumber\\
&\times\Omega(\mu_1,\lambda_1,i_{3},l_{3})\Omega(\nu_1,\kappa_1,j_{3},k_{3})\frac{\sqrt{i_{3}!j_3!k_3!l_3!}}{(\sqrt{2})^{i_3+j_{3}+k_{3}+l_{3}}}
			%\nonumber\\&\times
\delta_{i_{3}+l_{3},\mu_1+\lambda_1+i_2+l_2-\mu_2-\lambda_2}\delta_{j_{3}+k_{3},\nu_1+\kappa_1+j_2+k_2-\nu_2-\kappa_2}\nonumber\\
&\times\hat{d}_{1,\text{H}}^{\dagger i_1+l_1-\mu_1-\lambda_1}\hat{d}_{1,\text{V}}^{\dagger j_1+k_1-\nu_1-\kappa_1}
\hat{a}_{2,\text{H}}^{\dagger \mu_2+\lambda_2}\hat{a}_{2,\text{V}}^{\dagger \nu_2+\kappa_2}\left|\text{vac}\right\rangle,\nonumber\\\label{eq:Phip}
\end{align}
\end{widetext}
for $\bm{i}=(i_1i_2i_3)$, $\bm{j}=(j_1j_2j_3)$, $\bm{k}=(k_1k_2k_3)$ and~$\bm{l}=~(l_1l_2l_3)$
and 
\begin{align}
\label{eq:Omega}
	\Omega(\mu_1,\lambda_1,i_{3},l_{3})
		=&\sum_{\gamma=0}^{\mu_1+\lambda_1}{\mu_1+\lambda_1\choose \gamma}
			\nonumber\\&\times
		{{i_{3}+l_{3}-\mu_1-\lambda_1 }\choose {i_{3}-\gamma}}(-1)^{\mu_1+\lambda_1-\gamma}.
\end{align}
Eq.~(\ref{eq:Omega}) reduces to
\begin{align}
	\Omega&(\mu_1,\lambda_1,i_{3},l_{3})=(-1)^{\mu_1+\lambda_1}{{i_{3}+l_{3}-\mu_1-\lambda_1}\choose i_{3}}
			\nonumber\\&\times
		{}_2F_1(-\mu_1-\lambda_1,-i_{3};l_{3}-\mu_1-\lambda_1+1;-1)
\end{align}
for $ l_{3}-\mu_1-\lambda_1 \ge 0$ $\Omega$ and reduces to
 \begin{align}
	\Omega&(\mu_1,\lambda_1,i_{3},l_{3})=(-1)^{l_{3}}{{\mu_1+\lambda_1}\choose 
{\mu_1+\lambda_1- l_{3} }}
			\nonumber\\&\times
		{}_2F_1(-i_{3}-l_{3}+\mu_1+\lambda_1;l_{3};\mu_1+\lambda_1-l_{3}+1;-1)
\end{align}
for $ l_{3}-\mu_1-\lambda_1 < 0$,
and analogously for $\Omega(\nu_1,\kappa_1,j_{3},k_{3})$.

The quantum state after actual readout $\{\bm{q,r,s,t}\}$,
with $\bm{q}=(q_1q_2q_3)$ and similar for $\{\bm{r,s,t}\}$,
at the three fourtuples of detectors is
\begin{equation}
\rho=\prod_{u=1}^{3}\sum_{i_u,j_u,k_u,l_u=0}^{\infty}P^{\bm{qrst}}_{\bm{ijkl}}\left|
      {\Phi'}\right\rangle\left\langle{\Phi'}\right|
\label{eq:rho}
\end{equation}
with 
\begin{equation}
	P^{\bm{qrst}}_{\bm{ijkl}}
		=\frac{p(\bm{qrst}|\bm{ijkl})}
			{\prod\limits_{u=1}^{3}\sum\limits_{i_u,j_u,k_u,l_u=0}^{\infty}
				p\left(\bm{qrst}|\bm{ijkl}\right)\langle\Phi|\Phi\rangle}.
\label{eq:P}
\end{equation}
Entanglement verification can be done by measuring the coincidence rate, for various combinations of local projection measurements,
which in turn is done by introducing variable polarization rotators in the spatial paths of the modes~$d_1$ and~$a_2$
and then detecting them after passing through polarizing beam splitters as done in~\cite{SHS+09}. The polarization rotators are given by unitary operations
\begin{equation}
\hat{U}_{a_2}(\tilde{\alpha})=\exp\left[\frac{1}{2}i\tilde{\alpha}
      \left(\hat{a}^\dag_{2,V}\hat{a}_{2,H}+\hat{a}_{2,V}\hat{a}^\dag_{2,H}\right)\right]
\end{equation}
and
\begin{equation}
\hat{U}_{d_1}(\tilde{\delta})=\exp\left[\frac{1}{2}i\tilde{\delta}
       \left(\hat{d}^\dag_{1,V}\hat{d}_{1,H}+\hat{d}_{1,V}\hat{d}^\dag_{1,H}\right)\right].
\end{equation}

Given the imperfect Bell-state measurement events $\{\bm{q,r,s,t}\}$
on the three detector fourtuples,
the conditional probability that ideal measurements of  modes~$a_{2,H}$, $a_{2,V}$,$d_{1,V}$ and~$d_{1,H}$, would have yielded the result $\{i',j',k',l'\}$ is
\begin{align}
     p(i'j'k'l'|\bm{qrst})
 	=&\text{Tr}\{(|i'j'k'l'\rangle \langle i'j'k'l'|)\hat{U}_{a_2}(\tilde{\alpha})
			\nonumber\\&\otimes
		\hat{U}_{d_1}(\tilde{\delta}) \rho \hat{U}^\dag_{d_1}(\tilde{\delta}) \hat{U}^\dag_{a_2}(\tilde{\alpha})   
           \}\nonumber\\
    =&\prod_{u=1}^{3}\sum_{i_u,j_u,k_u,l_u=0}^{\infty}W^{\bm{ijkl}}_{i'j'k'l'}\times    
          P^{\bm{qrst}}_{\bm{ijkl}},
\end{align}
with~$P^{\bm{qrst}}_{\bm{ijkl}}$ given by Eq.~(\ref{eq:P}) and
\begin{align}
	W^{\bm{ijkl}}_{i'j'k'l'}
		:=&|\langle i'j'k'l'|\hat{U}_{a_N}(\tilde{\alpha}) \otimes\hat{U}_{d_1}(\tilde{\delta})|\Phi\rangle|^2
			\nonumber\\
		=&\left|A^{\bm{ijkl}}_{i'j'k'l'}\right|^2
\end{align}
is the transition probability.
Here
\begin{widetext}
\begin{align}
A^{\bm{ijkl}}_{i'j'k'l'}=&
    \prod_{p=1}^{2}\frac{1}{\sqrt{2^{i_p+j_p+k_p+l_p}i_p!j_p!k_p!l_p!}}\frac{(\tanh\chi)^{i_p+j_p+k_p+l_p}}       
         {\cosh^{4N}\chi}%\nonumber\\
   % &\times
        \sum_{\mu_p=0}^{i_p}\sum_{\nu_p=0}^{j_p}\sum_{\kappa_p=0}^{k_p}
         \sum_{\lambda_p=0}^{l_p}
            (-1)^{\mu_p+\nu_p}{i_p\choose \mu_p}{j_p\choose \nu_p}{k_p\choose \kappa_p}{l_p
                \choose \lambda_p}\nonumber\\
    &\times
        \Omega(\mu_1,\lambda_1,i_3,l_3)
            \Omega(\nu_1\kappa_1,j_3,k_3)%\nonumber\\
   % &\times
       \frac{\sqrt{i_3!j_3!k_3!l_3!}}       
        {(\sqrt{2})^{i_3+j_3+k_3+l_3}}%\nonumber\\
           \delta_{i_{3}+l_{3},         
               \mu_1+\lambda_1+i_2+l_2-\mu_2-\lambda_2}\delta_{j_{3}+k_{3},    
               \nu_1+\kappa_1+j_2+k_2-\nu_2-\kappa_2}\nonumber\\
        &\times(\nu_2+\kappa_2)!(j_1+k_1-\nu_1-\kappa_1)!\sqrt{\frac{j'!k'!}{i'!l'!}}
         \sum_{n_a=0}^{\text{Min}[j',\nu_2+\kappa_2]}
            \sum_{n_d=0}^{\text{Min}[k',j_1+k_1-\nu_1-\kappa_1]}%\nonumber\\
    %&\times
             (i\tan\frac{\tilde{\alpha}}{2})^{\nu_2+\kappa_2+j'-2n_a}(\cos\frac{\tilde{\alpha}}
           {2})^{i'+j'-2n_a}\nonumber\\
    &\times(i\tan\frac{\tilde{\delta}}{2})^{k'+j_1+k_1-\nu_1-\kappa_1-2n_d}
              (\cos\frac{\tilde{\delta}}{2})^{l'+k'-2n_d}%\nonumber\\
    %&\times
          \frac{(i'+j'-n_a)!(l'+k'-nd)!}{n_a!n_d!(j'-n_a)!(k'-n_d)!(\nu_2+\kappa_2-n_a)! 
           (j_1+k_1-\nu_1-\kappa_1-n_d)!}\nonumber\\
   &\times\delta_{i'+j',\mu_2+\nu_2+\kappa_2+\lambda_2}
\delta_{k'+l',i_1+j_1+k_1+l_1-\mu_1-\nu_1-\kappa_1-\lambda_1}.
\label{eq:AN=2}
\end{align}
\end{widetext}

The conditional probability to observe the event $\{q',r',s',t'\}$ on modes~$a_{2,H}$, $a_{2,V}$, $d_{1,V}$ and~$d_{1,H}$
with nonideal imperfect detectors, 
given imperfect Bell-state measurement events $\{\bm{q,r,s,t}\}$
at the three detector fourtuple is
\begin{align}
	Q:=&p(q'r's't'|\bm{qrst})\nonumber\\
	=&\sum_{i',j',k',l'=0}^{\infty}p(q'r's't'|i'j'k'l')p(i'j'k'l'|\bm{qrst}).
\label{eq:Q}
\end{align}
This equation is used to calculate the two-photon visibility.

\subsection{Reduction of states under ideal detectors}
\label{subsec:reduction}

In this subsection we consider the case of ideal detectors to show a reduction of the expressions
in the previous subsection to well known Bell state results. Our model incorporates transmission loss into the detector-efficiency parameter. Hence unit-efficiency detectection implies zero transmission loss.
For a single swap with imperfect threshold detectors, a non-ideal projection onto the Bell state $|\psi^-\rangle_{bc}$ is achieved whenever Bell-state measurement events $\{q,r,s,t\}=\{1,0,1,0\}$ or $\{0,1,0,1\}$ are obtained.

Let us consider the outcome $\{1,0,1,0\}$.
Thus, for ideal detectors with unit efficiency and zero dark counts,
\begin{equation}
	P_{ijkl}^{qrst}=\delta_{qi}\delta_{rj}\delta_{sk}\delta_{tl},
\end{equation}
and the state in Eq.~(\ref{eq:rho}) reduces to a single component.
For a single swap, Eq.~(\ref{eq:statePhi}) yields
\begin{equation}
	|\Phi_{1010}\rangle
		=\frac{1}{\sqrt 2}\left(\frac{|1010\rangle-|0101\rangle}{\sqrt 2}
			+\frac{|0011\rangle-|1100\rangle}{\sqrt 2}\right).\label{eq:Wi}
\end{equation}
Hence, there is another term superposed to a perfect four-mode singlet
\begin{equation}
	|\psi^-\rangle:=\frac{1}{\sqrt 2}\left(|1010\rangle-|0101\rangle\right).
\label{eq:singlet}
\end{equation}
The perfect Bell state $|\psi^-\rangle$ results from each source producing exactly one pair and the other two terms result when one source produces two pairs and the other source produces vaccuum. The probability for each of these three alternatives is proportional to $\chi^4$, which explains why the resultant state of remaining modes $a$ and $d$ in Eq.~(\ref{eq:Wi}) does not depend on $\chi$.

\begin{figure}
\includegraphics[scale=0.5]{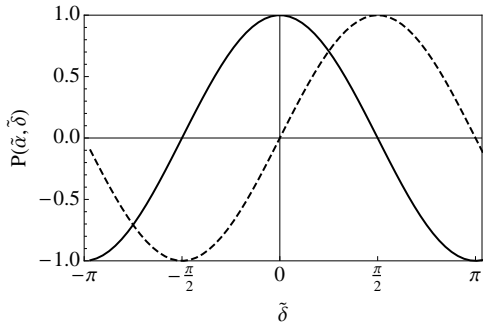}
\caption{
	Numerically evaluated correlation function $P(\tilde\alpha,\tilde\delta)$
	with $\tilde\alpha=0$ (solid curve) and $\tilde\alpha=\pi/2$ (dashed curve) for perfect detectors.
	Both curves are sinusoidal over $\tilde\alpha-\tilde\delta$.
     } 
\label{fig:PPerfect}
\end{figure}

For two concatenated elementary swaps, with the condition that all three detector fourtuples yield $\{1,0,1,0\}$,
the renormalized state $|\Phi'_{\bm{ijkl}}\rangle$ from Eq.~(\ref{eq:Phip}) yields
\begin{equation} 
	\left|\Phi'_{\bm{ijkl}}\right\rangle
		=\frac{1}{\sqrt 2}\left(\frac{|1010\rangle-|0101\rangle}{\sqrt 2}+\frac{|0011\rangle-|1100\rangle}{\sqrt 2}\right),
\end{equation}
which is the same state as Eq.~(\ref{eq:Wi}). Under these conditions, the conditional probabilities, $Q_{1010}$ and $Q_{0101}$, of recording the events $\{1,0,1,0\}$ and $\{0,1,0,1\}$, respectively, on modes~$a_{2,\text{H}}$, $a_{2,\text{V}}$, $d_{1,\text{V}}$ and $d_{1,\text{H}}$ are calculated from Eq.~(\ref{eq:Q}) as
\begin{align}
Q_{1010}=Q_{0101}
=A\cos^2\left(\frac{\tilde\alpha-\tilde\delta}{2}\right)
\label{eq:cos}
\end{align}
and the corresponding probabilities, $Q_{0110}$ and $Q_{1001}$ for the events $\{0,1,1,0\}$ and $\{1,0,0,1\}$ are
\begin{align}
Q_{1001}=Q_{0110}
=A\sin^2\left({\frac{\tilde\alpha-\tilde\delta}{2}}\right).
\label{sin}
\end{align}
Here, $A$ is the normalization factor.
The correlation function is
\begin{align}
	P(\tilde\alpha,\tilde\delta)
		:=&\frac{Q_{1010}+Q_{0101}-Q_{1001}-Q_{0110}}{Q_{1010}+Q_{0101}+Q_{1001}+Q_{0110}}\nonumber\\
=&\cos(\tilde\alpha-\tilde\delta),
\label{CorrelationFunction}
\end{align}
with $\tilde\alpha$ and $\tilde\delta$ characterizing the polarization rotators. We numerically simulate the two swaps case for ideal detectors with fixed $\tilde\alpha$ and varying $\tilde\delta$ with truncation at $n_\text{max}=1$. $P(\tilde\alpha,\tilde\delta)$ thus obtained is consistent with Eq.~(\ref{CorrelationFunction}) as shown in Fig.~\ref{fig:PPerfect}.

\subsection{Visibility for two concatenated elementary swaps}
\label{subsec:visibility}

\begin{figure}
	\includegraphics[scale=0.5]{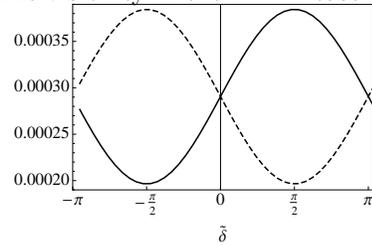}
\caption{
Four-fold coincidence probabilities $Q_{1001}+Q_{0110}$ (dashed curve) and $Q_{1010}+Q_{0101}$ (solid curve)
	plotted as a function of polarization rotation angle $\tilde\delta$ for fixed rotation angle $\tilde\alpha=\pi/2$. Detector parameters are dark count rate $\wp=1\times10^{-5}$ and efficiency $\eta=0.04$,
	and the pump parameter is $\chi=0.24$. The truncation is done at $n_\text{max}=3.$
} 
\label{fig:CoincidenceCount90}
\end{figure}

For imperfect detectors and losses, visibility decreases as the number of entanglement-swapping concatenations increases.
We calculate the visibility for the $N=2$ case from our model derived in Subsec.~\ref{subsec:conditional}.
For Bell-state measurement events $\{1,0,1,0\}$ or $\{0,1,0,1\}$ on all three detector fourtuples, 
which yields the singlet state~(\ref{eq:singlet}),
the two-fold coincidences $Q_{1010}+Q_{0101}$ and  $Q_{0110}+Q_{1001}$ are calculated numerically for dark count probability $\wp=1\times10^{-5}$, efficiency $\eta=0.04$ and source brightness $\chi=0.24$. The detector efficiency includes the channel loss given as $\eta=\eta_0\times10^{-\alpha l/10}$
 with $\alpha$ the loss coefficient, $l$ the distance that light travels and $\eta_0$ the intrinsic detector efficiency. As an example, for light with a wavelength of 1550~nm propagating through a telecom optical fibre, the loss coefficient is approximately $\alpha=0.2$~km$^{-1}$ and for  standard InGaAs avalanche photodiodes $\eta_0=0.15$. Thus  $\eta=0.04$ corresponds to a distance of 30~km in one arm. This corresponds to a total distance of about 240~km between left-most and right-most arm of the $ N=2$ setup. For superconducting detectors featuring $\eta=0.93$~\cite{MVS+13} these distances change to 70~km and 560~km respectively. The maximum number of photons in each mode are truncated at $n_\text{max}=3$.
The results of the numerical simulation are shown in Fig.~\ref{fig:CoincidenceCount90} for fixed angle~$\tilde\alpha=\pi/2$.
As expected, the two curves for $Q_{1010}+Q_{0101}$ and $Q_{1001}+Q_{0110}$ are complementary. Visibility is given as 
\begin{equation}
	V=\frac{\text{max}-\text{min}}{\text{max}+\text{min}}
\end{equation}
with~max denoting the maximum value of conditional two-fold coincidence
and~min the minimum value.
For our numerical simulations, the visibility is about $32\%$.
For the same detector parameters, loss and source brightness, the visibility for a single swap is about $70\%$. 
\begin{figure}
	\includegraphics[scale=0.5]{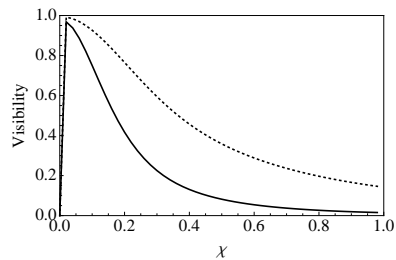}
\caption{
Comparison of the variation of visibility with source brightness $\chi$ for single swap, N=1(dotted curve) and for two elementary swaps, N=2 (solid curve). Both rotation angles are fixed at $\tilde\alpha=\tilde\delta=\pi/2$ and the detector parameters are the same as in Fig.~\ref{fig:CoincidenceCount90}. For the $N=2$ case, $n_\text{max}=3$ while for the $N=1$ case, $n_\text{max}=4$.
} 
\label{fig:N1N2Comparison}
\end{figure}
We now have expressions for the $N=2$ concatenated entanglement swapping case,
shown that they reduce to known results for perfect detectors
and numerically evaluated visibilities.
The visibility in the $N=2$ case is compared to that in the $N=1$ case in Fig.~\ref{fig:N1N2Comparison}. The visibility drops more rapidly for increasing $\chi$ for the $N=2$ case.
 
In the next section we develop the full formalism for the case of arbitrary~$N$.

\section{Concatenation of N swappings}
\label{sec:N}

\begin{figure}
	\includegraphics[scale=0.3]{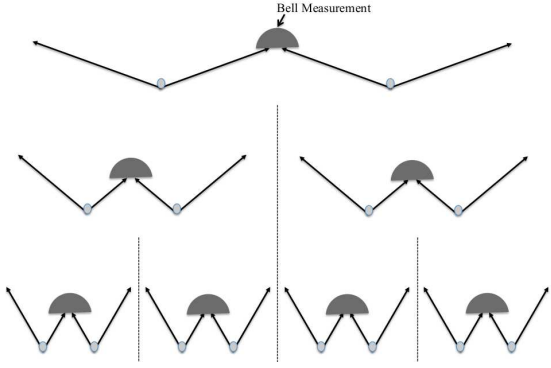}
\caption{
	Concatenation of $N=4$ elementary swaps:
        Filled-in caps are each a detector fourtuple and each element of the bottom row is an entanglement swapping resulting from a Bell measurement. The middle row depicts two $N=2$ cases corresponding to concatenating $N=1$ cases from the bottom row. The filled-in cap in the middle row corresponds to a detector 12-tuple. The top row corresponds to  $N=4$ which combines the two $N=2$ from the middle row. The filled-in cap in the top row corresponds to a detector 28-tuple.
     }
\label{fig:ConctES}
\end{figure}

We now extend the treatment of two concatenations to arbitrary distance and arbitrary number of swappings $N=2^\imath$ for $\imath$ any positive integer.
The 1$^\text{st}$ and the 2$^\text{nd}$
swaps are combined on the $(N+1)^\text{st}$ beam splitter,
and the $(N-1)^\text{st}$ and $N^\text{st}$ swaps combine on the $(N+\frac{N}{2})^\text{st}$ beam splitter etc.
Thus $2N-1$ Bell-state measurements are performed with $2N-1$ swaps. These swaps can be represented as an effective $N=1$ swapping with a detector $2^{N}-4$ tuple. In Fig.~\ref{fig:ConctES}, from 2nd row, there will be a total of $\imath$ rows and each $m$th row contains $N/2^m$ swaps. 
After these swappings the unnormalized state of remaining modes~$d_1$ and~$a_N$ is
\begin{widetext}
\begin{align}
\left|\Phi\right\rangle
        =&\prod_{p=1}^{N}\frac{1}
              {(\sqrt{2})^{i_p+j_p+k_p+l_p}\sqrt{i_p!j_p!k_p!l_p!}}\frac{(\tanh\chi)^{i_p+j_p+k_p+l_p}}
                {\cosh^{4N}\chi}%\nonumber\\
       %&\times
              \sum_{\mu_p=0}^{i_p}\sum_{\nu_p=0}^{j_p}\sum_{\kappa_p=0}^{k_p}
             \sum_{\lambda_p=0}^{l_p}(-1)^{\mu_p+\nu_p}{i_p\choose \mu_p}{j_p\choose 
                 \nu_p}
                 {k_p\choose \kappa_p}{l_p\choose \lambda_p}\nonumber\\
      &\times\prod_{m=1}^\imath\prod_{n=1}^{N/2^m}\Omega(\mu_{\beta_{mn}},
            \lambda_{\beta_{mn}},i_{\alpha_{mn}},l_{\alpha_{mn}})\Omega(\nu_{\beta_{mn}},
                \kappa_{\beta_{mn}},j_{\alpha_{mn}},k_{\alpha_{mn}})%\nonumber\\
     %&\times
           \frac{\sqrt{i_{\alpha_{mn}}!j_{\alpha_{mn}}!k_{\alpha_{mn}}!l_{\alpha_{mn}}!}}
            {(\sqrt{2})^{i_{\alpha_{mn}}+j_{\alpha_{mn}}+k_{\alpha_{mn}}+l_{\alpha_{mn}}}}\nonumber\\
        &    \times\delta_{i_{\alpha_{mn}}+l_{\alpha_{mn}},
           \mu_{\beta_{mn}}+\lambda_{\beta_{mn}}+i_{{\beta_{mn}}+1}+l_{{\beta_{mn}}+1}
              -\mu_{{\beta_{mn}}+1}-\lambda_{{\beta_{mn}}+1}}%\nonumber\\
     %&\times
             \delta_{j_{\alpha_{mn}}+k_{\alpha_{mn}},   
           \nu_{\beta{mn}}+\kappa_{\beta_{mn}}+j_{{\beta_{mn}}+1}+k_{{\beta_{mn}}+1}
              -\nu_{{\beta_{mn}}+1}-\kappa_{{\beta_{mn}}+1}}\nonumber\\
     &\times\hat{d}_{1,\text{H}}^{\dagger i_1+l_1-\mu_1-\lambda_1}
                    \hat{d}_{1,\text{V}}^{\dagger j_1+k_1-\nu_1-\kappa_1}
                       \hat{a}_{N,\text{H}}^{\dagger \mu_N+\lambda_N}
                         \hat{a}_{N,\text{V}}^{\dagger \nu_N+\kappa_N}\left|
                           \text{vac}\right\rangle.
\end{align}
\end{widetext}
Here,
\begin{equation}
	\alpha_{mn}=N\left(\frac{1-(\frac{1}{2})^m}{1-\frac{1}{2}}\right)+n \qquad \beta_{mn}=2^{m-1}(2n-1)
\end{equation}
whereas for $\l_\alpha-\mu_{\beta_{mn}}-\lambda_{\beta_{mn}}\ge 0$,
%\begin{widetext}
\begin{align}
	\Omega&(\mu_{\beta_{mn}},\lambda_{\beta_{mn}},i_{\alpha_{mn}},l_{\alpha_{mn}})\nonumber\\
     =&(-1)^{\mu_{\beta_{mn}}+\lambda_{\beta_{mn}}}{{i_{\alpha_{mn}}+l_{\alpha_{mn}}-\mu_{\beta_{mn}}-\lambda_{\beta_{mn}}}\choose i_{\alpha_{mn}}}\nonumber\\  
          &\times {}_2F_1(-\mu_{\beta_{mn}}-\lambda_{\beta_{mn}},-
               i_{\alpha_{mn}};l_{\alpha_{mn}}-\mu_{\beta_{mn}}-\lambda_{\beta_{mn}}+1;-1)
\end{align}
%\end{widetext}
 and for $\l_{\alpha_{mn}}-\mu_{\beta_{mn}}-\lambda_{\beta_{mn}}< 0$, 
\begin{widetext}
\begin{align}
  \Omega(\mu_{\beta_{mn}},
       \lambda_{\beta_{mn}},i_{\alpha_{mn}},l_{\alpha_{mn}})%\nonumber\\
   =&(-1)^{l_{\alpha_{mn}}}{{\mu_{\beta_{mn}}+\lambda_{\beta_{mn}}}\choose 
             {\mu_{\beta_{mn}}+\lambda_{\beta_{mn}}- l_{\alpha_{mn}} }}\nonumber\\
     &\times
         {}_2F_1(-i_{\alpha_{mn}}-
         l_{\alpha_{mn}}+\mu_{\beta_{mn}}+\lambda_{\beta_{mn}};l_{\alpha_{mn}};
         \mu_{\beta_{mn}}+\lambda_{\beta_{mn}}-l_{\alpha_{mn}}+1;-1).
\end{align}
\end{widetext}
The form of $\Omega(\nu_\beta,\kappa_\beta,j_\alpha,k_\alpha)$ is analogous.
The expression for $A^{\bm{ijkl}}_{i'j'k'l'}$ for this general case is calculated as
\begin{widetext}
\begin{align}
A^{\bm{ijkl}}_{i'j'k'l'}=&
    \prod_{p=1}^{N}\frac{1}{\sqrt{2^{i_p+j_p+k_p+l_p}i_p!j_p!k_p!l_p!}}\frac{(\tanh\chi)^{i_p+j_p+k_p+l_p}}       
         {\cosh^{4N}\chi}%\nonumber\\
   % &\times
        \sum_{\mu_p=0}^{i_p}\sum_{\nu_p=0}^{j_p}\sum_{\kappa_p=0}^{k_p}
         \sum_{\lambda_p=0}^{l_p}
            (-1)^{\mu_p+\nu_p}{i_p\choose \mu_p}{j_p\choose \nu_p}{k_p\choose \kappa_p}{l_p
                \choose \lambda_p}\nonumber\\
    &\times\prod_{m=1}^\imath\prod_{n=1}^{N/2^m}
        \Omega(\mu_{\beta_{mn}},\lambda_{\beta_{mn}},i_{\alpha_{mn}},l_{\alpha_{mn}})
            \Omega(\nu_{\beta_{mn}}\kappa_{\beta_{mn}},j_{\alpha_{mn}},k_{\alpha_{mn}})%\nonumber\\
   % &\times
       \frac{\sqrt{i_{\alpha_{mn}}!j_{\alpha_{mn}}!k_{\alpha_{mn}}!l_{\alpha_{mn}}!}}       
        {(\sqrt{2})^{i_{\alpha_{mn}}+j_{\alpha_{mn}}+k_{\alpha_{mn}}+l_{\alpha_{mn}}}}\nonumber\\
    &\times\delta_{i_{\alpha_{mn}}+l_{\alpha_{mn}},    
        \mu_{\beta{mn}}+\lambda_{\beta_{mn}}+i_{{\beta_{mn}}+1}+l_{{\beta_{mn}}+1}-
         \mu_{{\beta_{mn}}+1}-\lambda_{{\beta_{mn}}+1}}%\nonumber\\
    %&\times
         \delta_{j_{\alpha_{mn}}+k_{\alpha_{mn}},
          \nu_{\beta_{mn}}+\kappa_{\beta_{mn}}+j_{{\beta_{mn}}+1}+k_{{\beta_{mn}}+1}-
            \nu_{{\beta_{mn}}+1}-\kappa_{{\beta_{mn}}+1}}\nonumber\\
    &\times(\nu_N+\kappa_N)!(j_1+k_1-\nu_1-\kappa_1)!\sqrt{\frac{j'!k'!}{i'!l'!}}
         \sum_{n_a=0}^{\text{Min}[j',\nu_N+\kappa_N]}
            \sum_{n_d=0}^{\text{Min}[k',j_1+k_1-\nu_1-\kappa_1]}%\nonumber\\
    %&\times
             \left(i\tan\frac{\tilde{\alpha}}{2}\right)^{\nu_N+\kappa_N+j'-2n_a}\nonumber\\
    &\times\left(\cos\frac{\tilde{\alpha}}
           {2}\right)^{i'+j'-2n_a}\left(i\tan\frac{\tilde{\delta}}{2}\right)^{k'+j_1+k_1-\nu_1-\kappa_1-2n_d}
              \left(\cos\frac{\tilde{\delta}}{2}\right)^{l'+k'-2n_d}%\nonumber\\
    %&\times
          \nonumber\\
   &\times\frac{(i'+j'-n_a)!(l'+k'-nd)!}{n_a!n_d!(j'-n_a)!(k'-n_d)!(\nu_N+\kappa_N-n_a)! 
           (j_1+k_1-\nu_1-\kappa_1-n_d)!}\nonumber\\
    &\times\delta_{i'+j',
         \mu_N+\nu_N+\kappa_N+\lambda_N}
\delta_{k'+l',i_1+j_1+k_1+l_1-\mu_1-\nu_1-\kappa_1-\lambda_1}.
\label{eq:A}
\end{align}
\end{widetext}
The two-fold coincidences and hence visibility can be calculated using Eq.~(\ref{eq:Q}). The complexity of the algorithm to calculate the same scales as $n_\text{max}^{12N}$, where $N$ is the number of swaps and $n_\text{max}$ is the truncation of multiphoton incidences.

\section{Conclusions}
\label{sec:conclusions}

We have developed a theoretical framework for concatenated entanglement swapping for imperfect detectors, sources and loss. Our theory will be valuable in modelling a new generation of long-distance quantum communication
experiments based on the quantum relay as well as quantum repeaters, which both rely on concatenated swapping.
Our theory assumes that the number of entanglement swapping operations is of the form $N=2^\imath$
for $\imath$ a positive integer,
hence does not reduce to the previous practical entanglement swapping analysis of $N=1$~\cite{SHS+09}
for which $\imath=0$.
We develop the $N=2$ case extensively and show that it reduces to the known perfect-detector case,
and we solve numerically for a truncation of $n_\text{max}=3$ photons per mode.
 The truncation at $n_\text{max}=3$ is reliable for small values of $\chi$, including $\chi=0.24$, but fails to deliver correct results for larger values of  $\chi$.

Although the general case is exponentially expensive to solve in terms of $N$, which is proportional to the length of the channel, we are optimistic that further simplification can be found. 
The expressions are complicated nested products and sums of many terms,
but we have sampled those terms and find that most are negligibly small.
If a strategy can be found to eliminate all small terms,
then the computations could be performed for $N$ higher than $2$ as we report here.
Despite the current limitation of working with $N=2$ and not beyond,
the $N=2$ case is directly relevant to soon-to-be-realized concatenated entanglement-swapping experiments.

\acknowledgments
We acknowledge valuable discussions with Michael Lamoreux and Artur Scherer and financial support from AITF and NSERC. BCS appreciates Senior Fellow support from CIFAR. This research has been enabled by the use of computing resources provided by WestGrid and Compute/Calcul Canada.

\bibliography{References}
\bibliographystyle{unsrt}

\end{document}